\documentclass[sigconf, nonacm]{acmart}

\usepackage{booktabs}
\usepackage{amsmath}
\usepackage{algorithm}
\usepackage{algpseudocode}
\usepackage{multirow}

\begin{document}

\title{Synthetic Prefixes to Mitigate Bias in Real-Time Neural Query Autocomplete}


\author{Adithya Rajan}
\email{adithya.rajan@walmart.com}
\affiliation{%
  \institution{Walmart Global Technology}
  \city{Hoboken}
  \state{New Jersey}
  \country{USA}
}
\author{Xiaoyu Liu}
\email{xiaoyu.liu@walmart.com}
\affiliation{%
  \institution{Walmart Global Technology}
  \city{Hoboken}
  \state{New Jersey}
  \country{USA}
}
\author{Prateek Verma}
\email{prateek.verma@walmart.com}
\affiliation{%
  \institution{Walmart Global Technology}
  \city{Bellevue}
  \state{Washington}
  \country{USA}
}
\author{Vibhu Arora}
\email{vibhu.arora@walmart.com}
\affiliation{%
  \institution{Walmart Global Technology}
  \city{Sunnyvale}
  \state{California}
  \country{USA}
}
\renewcommand{\shortauthors}{Rajan et al.}
\begin{abstract}
  We introduce a data-centric approach for mitigating presentation bias in real-time neural query autocomplete systems through the use of synthetic prefixes. These prefixes are generated from complete user queries collected during regular search sessions where autocomplete was not active. This allows us to enrich the training data for learning to rank models with more diverse and less biased examples. This method addresses the inherent bias in engagement signals collected from live query autocomplete interactions, where model suggestions influence user behavior. Our neural ranker is optimized for real-time deployment under strict latency constraints and incorporates a rich set of features, including query popularity, seasonality, fuzzy match scores, and contextual signals such as department affinity, device type, and vertical alignment with previous user queries. To support efficient training, we introduce a task-specific simplification of the listwise loss, reducing computational complexity from O(n²) to O(n) by leveraging the query autocomplete structure of having only one ground-truth selection per prefix. Deployed in a large-scale e-commerce setting, our system demonstrates statistically significant improvements in user engagement, as measured by mean reciprocal rank and related metrics. Our findings show that synthetic prefixes not only improve generalization but also provide a scalable path toward bias mitigation in other low-latency ranking tasks, including related searches and query recommendations.
\end{abstract}


\keywords{learning to rank,
query autocomplete,
presentation bias,
synthetic data,
machine learning,
e commerce,
neural ranking}

\maketitle

\section{Introduction}
Query Autocomplete (QAC) is a fundamental feature of modern search systems, assisting millions of users daily by providing real-time suggestions that help complete queries as they type. By reducing typing effort and guiding users towards more popular or relevant queries, QAC not only improves user experience but also enhances the quality of downstream search results \cite{bar2011context} \cite{cai2016survey} \cite{mitra2015query}. Despite its widespread adoption, designing effective QAC systems remains challenging due to the need for low latency and high relevance in dynamic user contexts.
A typical QAC system operates in multiple stages. Initially, baseline scoring estimates the popularity of candidate query completions using historical frequency data. These candidates are then efficiently retrieved via indexing frameworks or prefix-based data structures. Finally, a re-ranking stage refines these suggestions by incorporating contextual signals like user behavior, session history, or device type \cite{cai2016survey}. Traditionally, this re-ranking is performed using linear scoring functions over a fixed feature set, prioritizing efficiency and simplicity.

However, linear models struggle to capture the complex, nonlinear interactions present in rich contextual features. As the feature space expands to include device information, query semantics, session context, and temporal factors, a linear hyperplane provides a limited approximation of the underlying function that predicts user preferences. Moreover, the representational capacity of linear models constrains their ability to benefit from large-scale data, limiting their overall effectiveness. To address these limitations, we adopt a neural learning-to-rank (LTR) model capable of modeling complex feature dependencies and delivering more accurate, context-aware rankings \cite{burges2010ranknet}\cite{pang2017neural}\cite{yuan2020deeppltr}.

Training neural LTR models demands substantial and diverse labeled data. While logs of QAC user engagement provide such data, they suffer from presentation bias: queries previously ranked higher are more likely to be clicked, resulting in a feedback loop that skews training data \cite{joachims2017unbiased} \cite{agarwal2010spatio}. To overcome this, we introduce a novel data augmentation strategy that synthesizes additional training samples by generating prefixes from full queries observed in general search sessions where autocomplete was not engaged. This augmentation diversifies the training distribution and mitigates presentation bias. 

Recognizing that synthetic data alone may not fully represent natural user behavior, since prefixes extracted from complete queries may not reflect actual typing patterns, we combine these synthetic samples with real QAC engagement data. This combined dataset balances the reduction of bias with the preservation of behavioral fidelity, enabling the neural ranker to produce suggestions that are both more contextually relevant and realistic. 

Given the strict latency requirements of real-time QAC systems, we train a shallow feed-forward neural network optimized using a listwise loss function. Listwise loss has been shown to outperform pointwise and pairwise objectives in ranking tasks by considering the entire candidate list during training \cite{xia2008listwise}. While listwise loss typically incurs $O(n^2)$ computational complexity due to comparison or permutation operations \cite{cao2007listwise}, the QAC setting offers a unique simplification: each interaction involves a single positive selection among candidates. Exploiting this, we approximate the listwise objective through pairwise loss computations with $O(n)$ complexity, significantly reducing training overhead without compromising ranking quality.
We deploy this neural LTR re-ranker in a large-scale, production QAC system handling millions of daily queries. Online A/B testing demonstrates that our model improves MRR by over 1 percent compared to the existing linear baseline, confirming the effectiveness of neural LTR models trained with balanced data and optimized loss functions for real-time search applications.

\section{Related Work}

Learning-to-Rank has been central to modern search, evolving from early gradient-boosted tree algorithms like RankSVM and LambdaMART used in industrial search systems. Neural-network-based LTR models such as RankNet, LambdaRank, and recent transformer architectures have further improved ranking quality by capturing rich feature interactions. Commercial platforms including LinkedIn \cite{wang2020efficient}, Taobao \cite{meng2025generative}, and Amazon \cite{singh2024dial} have successfully adopted neural LTR for QAC.

While LTR has been widely applied to document ranking, its use in QAC is more recent. Yuan and Kuang’s DeepPLTR model applies a context-aware neural pairwise LTR for autocomplete, demonstrating significant offline and online MRR improvements in e-commerce settings \cite{yuan2020deeppltr}. However, their approach still relies on pairwise loss and does not include any approach to mitigate presentation bias that gets introduced while leveraging QAC logs to train the model.

Training ranking models on click or QAC logs often introduces presentation bias: items ranked higher historically attract more clicks regardless of true relevance. Traditional solutions such as counterfactual LTR and propensity scoring adjust for bias through corrective weighting, but these methods can be complex and need strong assumptions \cite{joachims2017unbiased}. In this work, we tackle presentation bias from the training data perspective by augmenting training data via simulated prefix generation from general search logs, enabling a more diverse and unbiased candidate set. By combining these synthetic examples with actual QAC engagement data, we obtain a balanced training dataset that mitigates feedback loops while preserving behavioral realism.

Listwise losses are considered more aligned with ranking metrics such as NDCG and MRR compared to pointwise or pairwise objectives, yet they are typically computationally expensive, often requiring sorting or permutation-based scoring operations with complexity $O(n \log n)$ or worse. Recent research has explored approximations or smooth surrogates to reduce training overhead \cite{qin2010approximation}. In contrast, we identify a QAC-specific optimization: because each prefix event involves exactly one positive suggestion, we approximate the listwise objective using pairwise loss with $O(n)$ complexity, yielding efficiency gains without loss of accuracy.

In summary, while the Information Retrieval literature provides strong foundations in neural LTR, bias correction, and ranking loss design, our work is the first to integrate: (1) A scalable data augmentation strategy to reduce presentation bias, and (2) an efficient listwise-to-pairwise optimization for listwise ranking unique to real-time autocomplete scenarios, deployed and validated at production scale.

\section{Methodology}
\subsection{Problem Formulation}
We address the problem of context-aware ranking for QAC, where the goal is to predict and present a ranked list of full query suggestions given a user’s partial input (prefix) and surrounding contextual signals. In high-traffic, latency-sensitive environments such as e-commerce search, an effective QAC system must balance relevance, real-time inference constraints, and bias mitigation in learning from historical interaction data.
\subsection{Task Definition}
Formally, for a given user input prefix p, and a contextual vector c, the QAC system retrieves a candidate set of M query suggestions $\mathcal{Q}\ =\ \{q_1,\ q_2,\ \ldots,\ q_M\}$ from an indexed suggestion corpus. The task is to learn a scoring function $f\left(p,\ q,\ c;\ \theta\right)$ parameterized by $\theta$, such that the suggestions can be ranked by predicted relevance scores:
\begin{equation}
 \mathrm{Rank}\left(q_i\right)\ =\ \mathrm{argsort}\left(f\left(p,\ q_i,\ c;\ \theta\right)\right), \forall\ q_i\ \in\ \mathcal{Q}   
\end{equation}
The objective is to maximize the probability that the top-ranked suggestion matches the user’s actual selection, thereby improving MRR and other user engagement metrics.
\subsection{Candidate Generation}
Candidates are retrieved from a indexed storage system that supports both exact and fuzzy prefix matches. Given a prefix p, the system retrieves the top $M=50$ suggestions ranked by query-level popularity scores. Fuzzy matching allows the system to recover relevant completions even in the presence of misspellings or typographical errors, while an exact match feature is used downstream to prioritize more accurate suggestions during re-ranking. Our implementation of fuzzy matching retrieves documents which match the prefix to an edit distance of 1.
\subsection{Contextual Feature Representation}

Each candidate $q_i$ is represented using a rich feature vector capturing both static and dynamic properties:

\begin{itemize}
  \item \textbf{Query-level features:} Predicted query popularity, long-term seasonality trends (e.g., month-of-year features) \cite{verma2023seasonalitybasedrerankingecommerce}, and fuzzy match quality (binary or scalar).
  
  \item \textbf{Contextual signals:} These include:
  \begin{itemize}
    \item \textbf{Category alignment:} Measures whether the category of the current query suggestion matches the category of the user’s previous query within the same session. More precisely, the category of a query is determined by a classifier that maps queries to predefined nodes in a taxonomy, such as the department or vertical associated with the query. This signal helps prioritize suggestions that are contextually relevant to the user’s ongoing intent.
        
    \item \textbf{Device-level context:} Features derived from user device metadata.
  \end{itemize}
\end{itemize}

This combination enables the model to personalize suggestions based on both short-term behavior and long-term trends.

\subsection{Training Data and Labeling}
Training a robust LTR model for QAC requires large volumes of diverse, high-quality data that reflect real user preferences across a wide range of contexts. However, collecting such data from production systems introduces several challenges. Chief among them is presentation bias, where users are more likely to engage with top-ranked suggestions, creating a feedback loop that reinforces existing rankings \cite{joachims2017unbiased} \cite{ovaisi2020selectionbias}. To address this, we adopt a hybrid data generation strategy that combines real user interaction data from QAC sessions with synthetic data derived from general search query logs. While simulated data significantly improves coverage, it does not always capture the nuances of live user behavior. Specifically, engagement patterns observed in free-typed search queries may differ from those in autocomplete scenarios, where users are influenced by on-the-fly suggestions. 

To balance the realism of real QAC interactions with the diversity offered by synthetic events, we combine both data sources into a unified training set. The real QAC data provides high-fidelity signals that reflect actual user preferences and behavior, while the simulated data introduces a broader and less biased distribution of prefix-query pairs. We empirically tune the mixture ratio between the two to optimize generalization and offline ranking performance. This hybrid approach yields a scalable and refreshable training pipeline that supports the generation of millions of additional examples. It enables broader coverage across the prefix space and reduces dependence on suggestions already favored by the existing model. Our primary training signal comes from QAC engagement logs, where each interaction consists of a typed prefix, user context, a ranked list of suggestions, and a single selected query. The selected query is treated as the positive label, while all unclicked suggestions shown in the list are treated as explicit negatives as we assume these suggestions have user impression. Although this real-time data captures actual user behavior, it is inherently biased, since model suggestions influence user actions, limiting the diversity of prefix-query pairs observed.

To mitigate this bias, we supplement engagement data with synthetic training examples by simulating QAC interactions. We begin with full queries obtained from search logs where autocomplete was not active. These queries represent unbiased user intent, unaffected by prior model suggestions.

To simulate realistic prefix-query pairs, we first estimate a data-driven prefix length distribution $D(s)$, where $s$ is the character length of a full query from QAC logs. This distribution captures the most probable prefix length distribution at which users typically start engaging with autocomplete suggestions for query length $s$, allowing us to generate  possible prefixes given a user typed query.

For each  query in the simulated dataset,  $q \in Q_{\text{full}}$, we randomly simulate a prefix using distribution following $D(s)$.  The simulated  prefix is then submitted to the same QAC retrieval system used in production to retrieve a list of $M$ candidate suggestions. The original full query q is assigned as the positive sample, while all other retrieved suggestions serve as implicit negatives. For the same query which appears multiple times in the simulated dataset, multiple possible prefixes could be generated, and the frequency of each prefix will follow prefix length distribution $D(s)$. This approach introduces diverse and informative training signals that are not constrained by existing model outputs.

The algorithm below summarizes this process:
\begin{algorithm}
\caption{Simulated Prefix Generation}
\begin{algorithmic}[1]
\State Construct $D(s)$ from live QAC data:
  \State \hspace{1em} a. Record all (prefix, suggestion) engagement pairs
  \State \hspace{1em} b. Estimate prefix length distribution $D(s)$
\State Initialize training dataset $T \gets \emptyset$
\For{each query $q \in Q_{\text{full}}$}
    \State \hspace{1em} a. Simulate a prefix $p$ using $D(s)$
    \State \hspace{1em} b. Retrieve top $M$ suggestions $S = \{q_1, ..., q_M\}$
    \State \hspace{1em} c. Assign $q$ as the positive label
    \State \hspace{1em} d. Define negatives $N = S \setminus \{q\}$
    \State \hspace{1em} e. Add training instance $(p, q, N)$ to $T$
\EndFor
\State \Return $T$
\end{algorithmic}
\end{algorithm}

\subsection{Illustrative Example}
Consider the full query “black leather jacket” which appears in historical search logs but is absent from QAC session logs because it did not surface in autocomplete results during a specific time period. Clearly, the query is 20 characters long. Based on the estimated distribution $D(s)$, in one training sample, we simulate a prefix 'black l' and use the QAC retrieval system to obtain the top three suggestions: 'black leather jacket', 'black leather boots', and 'black leather gloves'. The original query 'black leather jacket' is treated as the positive sample, while the other two suggestions are used as negatives. In other training sample for the same query 'black leather jacket', other shorter or longer prefixes could be simulated, with the overall prefix length distribution following $D(s)$. This synthetic training instance is unaffected by model bias and helps improve generalization. 

\subsection{Learning Objective}
While listwise LTR objectives offer stronger alignment with ranking metrics like NDCG and MRR, they are computationally expensive. However, the QAC setting provides a unique optimization for any given prefix and context, there is only one positive selection by the user. As such, we approximate the listwise objective using a pairwise loss function over the observed positive and the M-1 unclicked negatives 
\begin{equation}
    \mathcal{L}=\sum_{\left(q^+,q^-\right)}l\left(f\left(p,q^+,c\right),f\left(p,q^-,c\right)\right)
\end{equation}
where $l$ is a standard pairwise ranking loss (e.g., hinge or logistic). This approximation reduces training complexity to $O(n)$ per event, allowing us to scale to large datasets while retaining the benefits of listwise training.

\subsection{Deployment Constraints}
Our model is implemented as a shallow feed-forward neural network to satisfy real-time latency constraints. It supports inference throughput of over 20000 requests per minute per pod each involving ranking 50 candidates. The architecture is designed to be compact and optimized for low-latency environments, making it suitable for deployment in large-scale production systems.

\subsection{Neural Network Architecture}
To meet the stringent latency requirements of real-time QAC systems, we design a shallow feed-forward neural network as the core ranking model \cite{pasumarthi2019tfranking}. While more expressive models such as deep neural networks or ensemble methods (e.g., XGBoost) offer lower training loss, their inference cost is often higher in production environments where the system must respond within a few milliseconds. In our experiments, we observed that even optimized implementations of XGBoost incur higher latency compared to a shallow neural network of comparable predictive power. Consequently, we prioritize a compact architecture that balances ranking performance and responsiveness.

\begin{figure*}[htbp]
  \centering
  \includegraphics[width=0.6\textwidth]{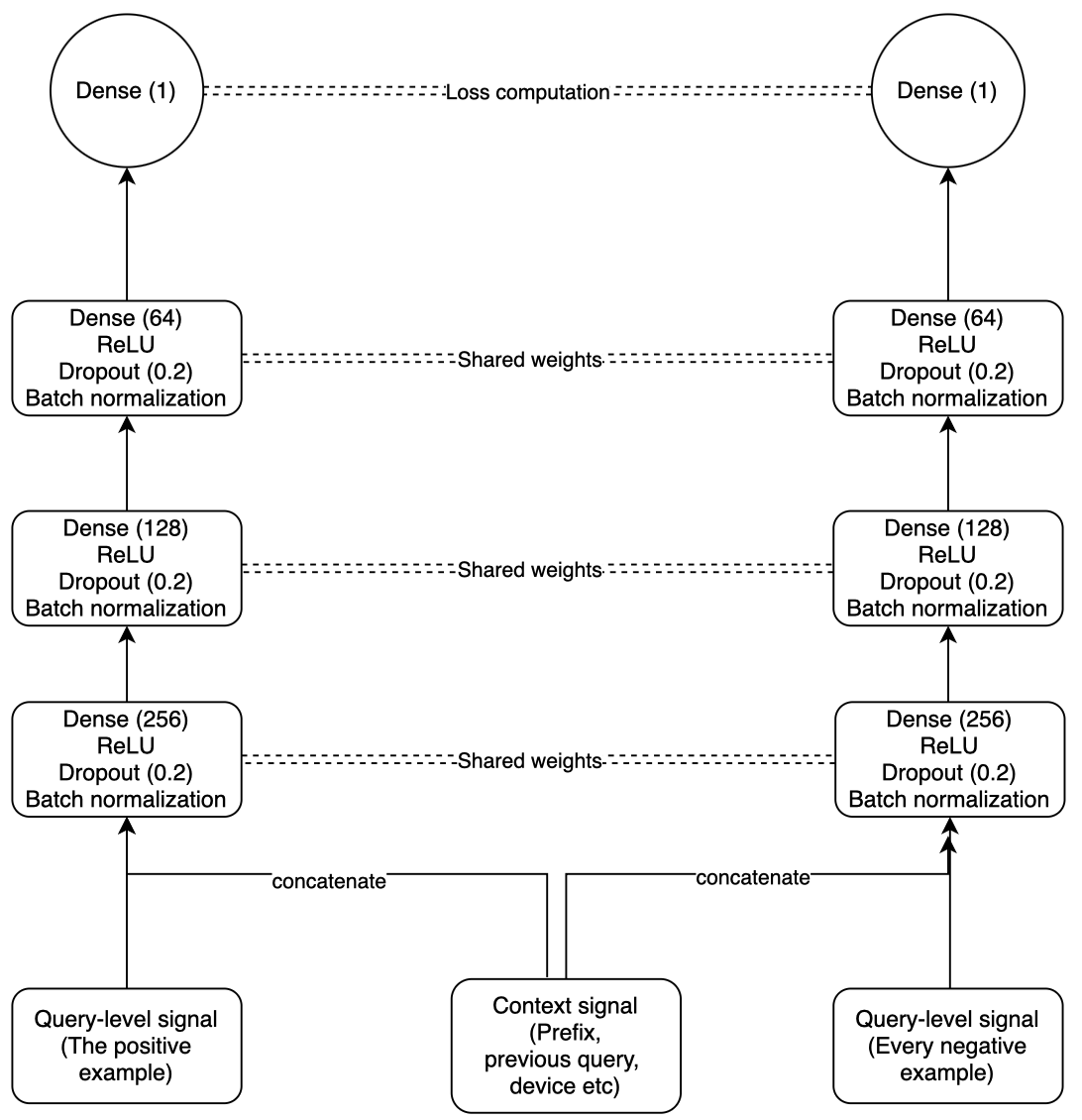}
  \caption{Model architecture of the Neural Ranker}
  \label{fig:neural_ranker}
\end{figure*}

\subsection{Model Structure}
The neural ranker is a fully connected feed-forward network with 3 hidden layers and 256, 128, 64 neurons per layer respectively as shown in the figure ~\ref{fig:neural_ranker}. Each layer uses the sigmoid activation function:
\begin{equation}
    h^{\left(l\right)}=\sigma\left(W^{\left(l\right)}h^{\left(l-1\right)}+b^{\left(l\right)}\right)
\end{equation}
where $h^{\left(l\right)}$ denotes the output of layer $l$, $W^{\left(l\right)}$ and $b^{\left(l\right)}$ are the weight matrix and bias vector for layer $l$, and $\sigma$ is the sigmoid function. The input layer $h^{\left(0\right)}$ is a concatenation of context and query-level features as shown in figure ~\ref{fig:neural_ranker}. Query-level signals are distinct for each query in the recall list while context signals are shared by all queries. Both of them consist of normalized scalar, boolean, and one-hot encoded categorical features. The final output layer produces a single scalar score per query suggestion:
\begin{equation}
\hat{y}=f\left(p,q,c;\theta\right)=W^{\left(out\right)}h^{\left(W\right)}+b^{\left(out\right)}
\end{equation}
This score is used to rank candidate suggestions for a given prefix and user context.
\subsection{Input Features}
Each query suggestion $q$ is represented as a fixed-length feature vector, consisting of:

\begin{itemize}
  \item \textbf{Query-level signals:}
  \begin{itemize}
    \item Predicted popularity score (normalized scalar)
    \item Long-term seasonality indicator (e.g., month-of-year) \cite{verma2023seasonalitybasedrerankingecommerce}
    \item Fuzzy match flag (boolean: whether the query starts with the prefix, or contains the prefix)
    \item Query department category (one-hot encoded), query vertical category (one-hot encoded)
    \item Number of tokens in the query
    \item Query length (number of characters)
  \end{itemize}

  \item \textbf{Contextual signals:}
  \begin{itemize}
    \item Department category match with previous query (Boolean)
    \item Vertical category match with previous query (Boolean)
    \item Device type (one-hot vector encoding: iOS app, Android app, desktop browser, mobile browser, etc.)
    \item Prefix length (number of characters)
    \item Number of tokens in prefix
  \end{itemize}
\end{itemize}

No learned embeddings are used in this architecture. All categorical and boolean variables are either one-hot encoded or directly fed into the model as binary indicators. Scalar features are standardized prior to training to improve convergence.

\subsection{Training Configuration}
The model is trained using the Adam optimizer with a batch size of 1280.  We apply dropout regularization between layers and include L2 weight decay to prevent overfitting. The training objective is a listwise loss (as described in Section 3), where each event comprises one positive suggestion and multiple negative candidates sampled from the unclicked QAC suggestions presented to the user.

\subsection{Inference Performance}
The model is optimized for high-throughput, low-latency inference. It is deployed in a real-time system that handles over tens of thousands of QAC requests per second, each involving the re-ranking of 50 candidate queries. Under production load, our model’s performance is well within the latency thresholds of any reasonable real-time system. This makes the architecture suitable for real-world, user-facing search applications where speed and relevance are both critical.

\section{Experimental Setup}
To evaluate the effectiveness of our proposed neural LTR approach for QAC, we conduct both offline and online experiments designed to assess ranking quality, user engagement, and system performance under production constraints. Our training and evaluation data is derived from 30 days of search logs and QAC engagement logs from a large-scale e-commerce platform. From these logs, we construct a dataset of 12 million QAC events, sampled uniformly across device types to ensure broad generalization. We split the data into 80 percent for training and 20 percent for testing. To assess the impact of simulated data augmentation, we experiment with three training set configurations: (i) 100 percent real QAC data, (ii) a balanced 50-50 mix of simulated and real QAC data, and (iii) 100 percent simulated data.

Offline evaluation is performed using two complementary metrics: $MRR_{QAC}$ and $MRR_{general}$. $MRR_{QAC}$ is computed over held-out QAC engagement logs and reflects the degree to which the model improves ranking quality relative to the previous production system. However, this metric is inherently influenced by presentation bias, since users disproportionately interact with suggestions shown at higher ranks. To mitigate this bias and evaluate generalization, we compute $MRR_{general}$ over query completions derived from general search logs, which may or may not originate from QAC usage. While $MRR_{general}$ is less biased, it has its own limitations: many of the prefixes used in this evaluation may not correspond to realistic QAC usage patterns, and the observed completions may reflect user behaviors that bypass QAC entirely (e.g., typing out short queries or pasting long ones). Thus, these two metrics provide complementary views of model quality: $MRR_{QAC}$ reflects alignment with existing user engagement, while $MRR_{general}$ probes potential for improved relevance across a wider spectrum of queries.

Online evaluation is conducted through a large-scale A/B test, where the control group is served by the existing linear ranker, and the treatment group is served by our proposed neural LTR model. Both groups share identical indexes and retrieval logic to ensure that differences arise solely from re-ranking. Each group receives an equal share of production traffic. We evaluate online performance using MRR, QAC usage rate, and guardrail metrics such as downstream click-through rate (CTR) and conversion rate. These ensure that improvements in QAC relevance do not negatively affect overall user engagement with search.

Through this rigorous experimental setup, we aim to assess the tradeoffs between different data augmentation strategies, the effectiveness of the neural ranker under real-time constraints, and its impact on user behavior at scale.

\section{Results}
We evaluate the performance of our neural LTR model through both offline experimentation and large-scale online A/B test. Our primary objective is to assess whether the model improves ranking quality under real-time latency constraints, while also examining the impact of different training data mixtures and user behavior patterns.

In offline evaluations, we compare the neural ranker against a production baseline trained using a linear scoring function. We report results using two distinct metrics: $MRR_{QAC}$, which measures mean reciprocal rank on historical QAC engagement data, and $MRR_{general}$, which evaluates ranking quality using prefix-query pairs derived from general search logs. The former provides a reliable proxy for in-situ user relevance but is known to suffer from presentation bias. The latter offers a broader view of potential generalization, although it includes prefixes that may not reflect real QAC usage patterns.

We experimented with three data construction strategies: a model trained purely on real QAC engagement data, one trained entirely on simulated prefix-query completions from general search logs, and a third trained on a 50-50 mix of the two. Models trained exclusively on real QAC data achieved the highest $MRR_{QAC}$ scores, indicating strong alignment with actual user interactions in the live system. Models trained solely on simulated data performed best on $MRR_{general}$, suggesting improved relevance on long-tail queries and underrepresented prefixes. However, this gain came at the cost of degraded $MRR_{QAC}$, likely due to misalignment with natural user behavior. The model trained on a 50-50 mix achieved moderate improvements across both metrics, offering a promising balance between realism and generalization. We briefly outline these offline metrics to establish the superiority of the augmented mix in the context of mitigating presentation bias as well as providing better rankings on real data in table ~\ref{tab:mrr_results}.

\begin{table}[h]
\centering
\renewcommand{\arraystretch}{1.3}
\begin{tabular}{lcc}
\toprule
\textbf{Training Strategy} & $\Delta\text{MRR}_{\text{QAC}}$ & $\Delta \text{MRR}_{\text{general}}$ \\
\midrule
QAC training data only       & +1.5\% & -1.6\% \\
Simulated prefix training data only & -2.7\% & +1.2\% \\
50-50 mix                    & +0.6\% & +0.2\% \\
\bottomrule
\end{tabular}
\caption{Offline evaluation results for different training strategies.}
\label{tab:mrr_results}
\end{table}

We conducted additional analysis of MRR improvements using a 50/50 blend of real QAC data and simulated training data across platforms, namely desktop, mobile web, iOS, and Android in order to showcase that user behavior varies by device. We also evaluated the impact of the neural ranker in sessions where a previous query was present and could inform the current ranking. The corresponding metrics are presented in table ~\ref{tab:platform_context_metrics}.

\begin{table}[ht]
\centering
\begin{tabular}{lll}
\toprule
\textbf{Platform / Context} & $\Delta\text{MRR}_{\text{QAC}}$ & $\Delta \text{MRR}_{\text{general}}$ \\
\midrule
\textbf{Platform} & \\
Desktop & +0.98\% & +0.79\%\\
Mobile Web & +3.29\% & +1.16\%\\
iOS & +0.62\% & +0.37\%\\
Android & +1.32\% & +0.19\%\\
\midrule
\textbf{Context} & \\
without previous\_query & +0.97\% & +0.07\%\\
with previous\_query & +1.22\% & +0.7\%\\
\bottomrule
\end{tabular}
\caption{Offline evaluation by Platform and Context}
\label{tab:platform_context_metrics}
\end{table}

\begin{table}[t]
\centering
\renewcommand{\arraystretch}{1.3}
\begin{tabular}{lcccc}
\toprule
 \textbf{} & \textbf{iOS} & \textbf{Android} & \textbf{dWeb} & \textbf{mWeb} \\
\midrule
MRR & 0.92\% & 0.79\% & 1.49\% & 1.25\% \\
\bottomrule
\end{tabular}
\vspace{6pt}
\caption{Online evaluation results across platforms}
\label{tab:platform_metrics}
\end{table}

In our online A/B test, we deploy the neural ranker alongside the existing linear ranker, splitting production traffic evenly between the two while ensuring identical retrieval backends and candidate sets. The neural ranker yielded a statistically significant increase of +0.92\% to 1.49\% in mean reciprocal rank relative to the linear baseline across platforms. Additionally, we observed an improvement of 0.89\% to 1.40\% in average query click position across platforms, indicating that relevant suggestions were surfaced earlier in the ranked list. Platform-level analysis revealed a +0.36\% increase in QAC usage for Android devices. Guardrail metrics, including overall clicks per search and conversion rates, remained neutral, affirming that the model’s gains in QAC ranking did not negatively impact downstream user engagement.

Inference latency was another critical metric, given the stringent constraints of a real-time QAC system. While the linear ranker operates with sub-millisecond p99 latency, the neural model maintained a p99 latency of double digit milliseconds. This remained well within the system's latency threshold and demonstrated the feasibility of deploying a neural architecture without sacrificing responsiveness.

These results collectively indicate that the neural LTR ranker improves QAC ranking effectiveness in both offline and online settings, especially when trained on a balanced mix of real and simulated data. The model offers consistent performance across device types and supports low-latency inference, making it suitable for production-scale autocomplete applications.
\section{Conclusion and Future Work}
In this paper, we present a novel approach to training and deploying a neural learning-to-rank model for query autocomplete systems under stringent latency constraints. Central to our contribution is a data augmentation method that effectively combines real QAC engagement data with simulated prefix-query pairs derived from general search logs. The ratio of real QAC engagement data to simulated data can be determined algorithmically or empirically. This hybrid training dataset helps mitigate presentation bias inherent in engagement logs while preserving alignment with actual user behavior. Additionally, we optimize the training process by leveraging pairwise loss computations within a listwise learning objective, significantly reducing computational overhead without sacrificing ranking quality. Our extensive offline and online evaluations demonstrate that the proposed neural ranker substantially outperforms the existing linear ranker, resulting in statistically significant improvements in mean reciprocal rank and user engagement metrics.

Our findings also reinforce the limitations of linear ranking functions in modeling the complex, nonlinear relationships present in user interaction data. While linear models offer simplicity and low latency, they fall short in capturing the rich contextual signals necessary for optimal ranking performance in modern QAC systems.

We acknowledge that the adoption of a neural LTR model introduces increased training complexity and computational cost. The integration of additional ranking signals requires careful experimentation to account for feature interactions and avoid degradation in performance. Future work will focus on enhancing context modeling by incorporating richer session-level features and refining device-specific adaptations to further personalize query suggestions. We also plan to explore deeper neural architectures and embedding-based features, balanced with latency requirements, to push the boundaries of QAC relevance and responsiveness.

Overall, this work lays a strong foundation for applying advanced neural ranking techniques to real-time query suggestion systems with promising avenues for continued improvement and broader applicability.

\section*{Acknowledgement}
We would like to thank our Platform Engineering team for their invaluable help and support—especially Mayank Lara, Dagshayani Kamalaharan, Kevin Li, and Sanjay Shah. We are also grateful to our product partners, Fawn Qiu and Keshav Agrawal, as well as our leadership, John Yan and Michael Bowersox, for their guidance and support throughout this work.

\bibliographystyle{unsrt}
\bibliography{references}

\end{document}